\documentclass[aps,prl,twocolumn,numerical,superscriptaddress,nofootinbib,showpacs,showkeys,longbibliography]{revtex4-1}

\usepackage{graphicx,color}
\usepackage{amsmath,amssymb,amsfonts}

\newcommand{\be}{\begin{equation}}
\newcommand{\ee}{\end{equation}}
\newcommand{\ba}{\begin{eqnarray}}
\newcommand{\ea}{\end{eqnarray}}
\newcommand{\ban}{\begin{eqnarray*}}
\newcommand{\ean}{\end{eqnarray*}}

\def\v2{\mbox{$v_2$}}

\def\sqrtsNN{\mbox{$\sqrt{s_{\mathrm{NN}}}$}}

\begin{document}

\title{A sensitivity study of the primary correlators used to characterize \\ chiral-magnetically-driven 
charge separation}
\medskip

\author{Niseem~Magdy} 
\email{niseemm@gmail.com}
\affiliation{Department of Physics, University of Illinois at Chicago, Chicago, Illinois 60607, USA}

\author{Mao-Wu Nie}
\affiliation{Institute of Frontier and Interdisciplinary Science, Shandong University, Qingdao, Shandong, 266237, China}
\affiliation{Key Laboratory of Particle Physics and Particle Irradiation, Ministry of Education, Shandong University, 
Qingdao, Shandong, 266237, China}

\author{Guo-Liang Ma}
\email[]{glma@fudan.edu.cn}
\affiliation{Key Laboratory of Nuclear Physics and Ion-beam Application (MOE), Institute of Modern Physics, Fudan University, Shanghai 200433, China}
\affiliation{Shanghai Institute of Applied Physics, Chinese Academy of Sciences, Shanghai 201800, China}

%
%

\author{Roy~A.~Lacey} 
\email{Roy.Lacey@stonybrook.edu}
\affiliation{Depts. of Chemistry \& Physics, Stony Brook University, Stony Brook, New York 11794, USA}

\date{\today}

\begin{abstract}
 A Multi-Phase Transport (AMPT)  model is used to study the detection sensitivity of two of the primary correlators 
-- $\Delta\gamma$ and $R_{\Psi_{2}}$ -- employed to characterize charge separation induced by 
the Chiral Magnetic Effect (CME). The study, performed relative to several event planes for different 
input ``CME signals'', indicates a detection threshold for the fraction 
$f_{\rm CME}=\Delta\gamma_{\rm CME}/\Delta\gamma$, which renders 
the $\Delta\gamma$-correlator insensitive to values of the Fourier dipole coefficient  $a_1 \alt 2.5\%$, 
that is larger than the purported signal(signal difference) for ion-ion(isobaric) collisions.
By contrast, the $R_{\Psi_{2}}$ correlator indicates concave-shaped distributions with inverse 
widths ($\mathrm{\sigma^{-1}_{R_{\Psi_2}}}$) that are linearly 
proportional to $a_1$, and independent of the character of the event plane used  for their extraction. 
The sensitivity of the $R_{\Psi_{2}}$ correlator to minimal CME-driven charge separation in the 
presence of realistic backgrounds, could aid better characterization of the CME in heavy-ion collisions.
\end{abstract}

\pacs{25.75.-q, 25.75.Gz, 25.75.Ld}
\maketitle

Ion-Ion collisions at both the Relativistic Heavy Ion Collider (RHIC) and the Large Hadron Collider (LHC) create 
hot expanding fireballs of quark-gluon plasma (QGP) in the background of a strong magnetic 
field \cite{Skokov:2009qp,McLerran:2013hla,Tuchin:2014iua}. 
Topologically nontrivial sphaleron transitions [via the axial anomaly] ~\cite{Manton:1983nd,Klinkhamer:1984di,Moore:2000ara} 
can induce different densities of right- and left-handed quarks in the plasma fireballs, resulting in a quark electric current 
along the $\vec{B}$-field. This phenomenon of the generation of a quark electric current ($\vec{J}_Q$) in the presence of a magnetic field is termed  the chiral magnetic effect (CME) \cite{Kharzeev:2004ey,Fukushima:2008xe}:
\begin{eqnarray} \label{eq_cme}
\vec{J}_Q &=& \sigma_5 \vec{B},\ \ \ \ \sigma_5 = \mu_5 \frac{Q^2}{4\pi^2},
\end{eqnarray} 
where, $\sigma_5$ is the chiral magnetic conductivity, 
$\mu_5$ is the chiral chemical potential that quantifies the axial charge 
asymmetry or imbalance between right- and left-handed quarks in the plasma, and $Q$ is the  electric 
charge \cite{Fukushima:2008xe,Son:2009tf,Zakharov:2012vv,Fukushima:2012vr}. 

Full characterization of the CME, which manifests experimentally 
as the separation of electrical charges along the $\vec{B}$-field~\cite{Kharzeev:2004ey,Fukushima:2008xe},  can give 
fundamental insight on anomalous transport and the interplay of chiral symmetry restoration, 
axial anomaly and gluon topology in the QGP 
\cite{Moore:2010jd,Mace:2016svc,Liao:2010nv,Kharzeev:2015znc,Skokov:2016yrj}. 

Charge separation stems from the fact that the CME preferentially drives charged particles, 
originating from the same ``P-odd domain'', along or opposite to the $\vec{B}$-field 
depending on their charge. This separation can be quantified via measurements of the first $P$-odd 
sine term ${a_{1}}$, in the Fourier decomposition of the charged-particle azimuthal 
distribution~\cite{Voloshin:2004vk}:
\begin{eqnarray}\label{eq:a1}
{\frac{dN^{\rm ch}}{d\phi} \propto [1 + 2\sum_{n} v_{n} \cos(n \Delta\phi) + a_n \sin(n \Delta\phi)  + ...]}
\end{eqnarray}
where $\mathrm{\Delta\phi = \phi -\Psi_{RP}}$ gives the particle azimuthal angle
with respect to the reaction plane (${\rm RP}$) angle, and ${v_{n}}$ and ${a_{n}}$ denote the
coefficients of $P$-even and $P$-odd Fourier terms, respectively. 
A direct measurement of the P-odd coefficients  $a_n$, is not possible due to the 
strict global $\cal{P}$ and $\cal{CP}$ symmetry of QCD.
However, their fluctuation and/or variance $\tilde{a}_n= \left<a_n^2 \right>^{1/2}$ can 
be measured with suitable correlators.

The CME-driven charge separation is small because only a few particles from the same $P$-odd domain 
are correlated.  Moreover, both the initial axial charge and the time evolution of the magnetic field (c.f. Eq.~\ref{eq_cme}) are 
unconstrained theoretically, and it is uncertain whether an initial CME-driven charge separation could survive 
the signal-reducing effects of the reaction dynamics, and still produce a signal above the detection threshold. 
Besides, it is uncertain whether a charge separation that survives the expansion dynamics would still be 
discernible in the presence of the well-known background correlations which contribute and complicate 
the measurement of CME-driven charge separation
\cite{Wang:2009kd,Bzdak:2010fd,Schlichting:2010qia,Muller:2010jd,Liao:2010nv,Khachatryan:2016got}.
Thus, the correlators used to characterize the CME, not only need to suppress background-driven 
charge-dependent correlations, such as the ones from resonance decays, charge ordering in jets, etc.,
but should also be sensitive to small charge separation signals in the presence of these backgrounds.
The latter requirement is  especially important for ongoing measurements [at RHIC] designed to detect 
the small signal difference between the Ru+Ru and Zr+Zr isobars~\cite{Adam:2019fbq}.

In this work we use the AMPT model  \cite{Lin:2004en} with varying amounts of input charge separation $\Delta S$, characterized 
by the partonic dipole term $a_1$, to study the detection sensitivity of the $\Delta\gamma$ and the $R_{\Psi_2}(\Delta S)$ correlators. 
The model is known to give a good representation of the experimentally measured 
particle yields, spectra, flow, etc.,\cite{Lin:2004en,Ma:2016fve,Ma:2013gga,Ma:2013uqa,Bzdak:2014dia,Nie:2018xog}. 
Therefore, it provides a realistic estimate of both the magnitude and the properties of the background-driven charge 
separation one might encounter in the data sets collected at RHIC and the LHC.

For these sensitivity tests, we simulated Au+Au collisions at $\sqrtsNN~=~200$ GeV with the 
new version of the AMPT model that incorporates string melting and local charge conservation.
There are four primary ingredients for each of these collisions:  (i) an initial-state, (ii) a parton cascade phase, 
(iii) a hadronization phase in which partons are converted to hadrons, and 
(iv) a hadronic re-scattering phase prior to kinetic freeze-out. 
The initial-state mainly simulates the spatial and momentum distributions of minijet partons from QCD 
hard processes and soft string excitations as encoded in the HIJING model~\cite{Wang:1991hta,Gyulassy:1994ew}. 
The parton cascade takes account of the strong interactions among partons through elastic partonic collisions 
controlled by a parton interaction cross section~\cite{Zhang:1997ej}. Hadronization, or the conversion from partonic 
to hadronic matter, is simulated via a coalescence mechanism. Subsequent to hadronization, the ART model is 
used to simulate baryon-baryon, baryon-meson and meson-meson 
interactions~\cite{Li:1995pra}.  

 A formal mechanism for the CME is not implemented in AMPT. However, modifications can be made to 
the model to mimic CME-induced charge separation~\cite{Ma:2011uma} by switching the $p_y$ values of a 
fraction of the downward moving $u$ ($\bar{d}$) 
quarks with those of the upward moving $\bar{u}$ ($d$) quarks to produce a net charge-dipole separation in 
the initial-state. Here, the $x$ axis is along the direction of the impact parameter $b$, the $z$ axis points along the 
beam direction, and the $y$ axis is perpendicular to the $x$ and $z$ directions, {\em i.e}, the direction of the 
proxy $\vec{B}$-field. 
The strength of the proxy CME signal is regulated by the fraction $f_{0}$ of the 
initial input charge separation \cite{Ma:2011uma,Huang:2019vfy}: 
\begin{equation}
f_{0} = \frac{N_{\uparrow(\downarrow)}^{+(-)}-N_{\downarrow(\uparrow)}^{+(-)}}{N_{\uparrow(\downarrow)}^{+(-)}+N_{\downarrow(\uparrow)}^{+(-)}},
\quad f_{0} =\frac{4}{\pi}a_{1}
 \label{eq-f}
\end{equation}
where $N$ is the number of a given species of quarks, $``+''$ and $``-''$ denote positive and negative charges, respectively, 
and $\uparrow$ and $\downarrow$ represent the directions along and opposite to that of the $y$ axis. 
Eq.~\ref{eq-f} also shows that the fraction $f_0$, is related to the $P$-odd dipole term $a_{1}$, defined in Eq.~\ref{eq:a1}. 
Note that this initial partonic charge separation $a_1$, is different from the final hadrons' charge 
separation $a_1$, often referred to in the literature and implemented in other models. 
Cross-checks made with the Anomalous-Viscous Fluid Dynamics model~\cite{Shi:2017cpu,Jiang:2016wve}
suggests that the two are linearly proportional to a very good approximation.
Simulated events, generated for a broad set of $f_0$ values, were analyzed with both the 
$\Delta\gamma$ and the $R_{\Psi_2}(\Delta S)$ correlators, to evaluate their respective sensitivity 
as discussed below quantitatively.

The charge-dependent correlator, $\gamma_{\alpha\beta}$~\cite{Voloshin:2004vk} ,  has been widely used 
at  RHIC \cite{Abelev:2009ac,Abelev:2009ad,Adamczyk:2013hsi,Adamczyk:2013kcb,
Adamczyk:2014mzf,Tribedy:2017hwn,Zhao:2017nfq}  
and the LHC \cite{Abelev:2012pa,Khachatryan:2016got} in ongoing attempts to 
identify and quantify CME-driven charge separation:
\ba
\
\gamma_{\alpha\beta} =& \left\langle \cos\big(\phi_\alpha^{(\pm)} +
\phi_\beta^{(\pm)} -2 \Psi_{\rm 2}\big) \right\rangle, \nonumber  \quad
\Delta\gamma =& \gamma_{\beta} - \gamma_{\alpha},
\label{eq:2}
\ea
where $\phi_\alpha,\phi_\beta$ denote the azimuthal emission angles for like-sign   ($++,\,--$)
and unlike-sign ($+\,-$) particle pairs. 
The question as to whether the experimental measurements for $\Delta\gamma$ indicate the CME, 
remain inconclusive because of several known sources of background correlations that can account for most, if not all, of the 
measurements~\cite{Wang:2009kd,Bzdak:2010fd,Schlichting:2010qia,Muller:2010jd,Liao:2010nv}.

A recent embellishment to the $\Delta\gamma$ correlator is the proposal to leverage 
the ratios of $\Delta\gamma$ and elliptic flow ($v_2$) measurements,  obtained 
relative to the reaction plane ($\Psi_{\rm RP}$) and the participant plane ($\Psi_{\rm PP}$)
\ba
  r_1 = \frac{\Delta\gamma(\Psi_{\rm RP}) }{\Delta\gamma(\Psi_{\rm PP}) }, 
	\quad r_2 = \frac{v_2(\Psi_{\rm RP}) }{v_2(\Psi_{\rm PP}) },
	\label{eq:3a}
\ea
to simultaneously constrain the CME and  background (${\rm Bkg}$) contributions 
to $\Delta\gamma$ \cite{Xu:2017qfs,Voloshin:2018qsm}:
\ba
\Delta\gamma(\Psi_{\rm PP}) = \Delta\gamma_{\rm CME}(\Psi_{\rm PP}) + \Delta\gamma_{\rm  Bkg}(\Psi_{\rm PP}),  \nonumber \\
\Delta\gamma(\Psi_{\rm RP}) = \Delta\gamma_{\rm CME}(\Psi_{\rm RP}) + \Delta\gamma_{\rm Bkg}(\Psi_{\rm RP}),   
\ea
and
\ba
\Delta\gamma_{\rm CME}(\Psi_{\rm PP}) = r_2 \times \Delta\gamma_{\rm CME}(\Psi_{\rm RP}), \nonumber \\
\Delta\gamma_{\rm Bkg}(\Psi_{\rm RP}) = r_2 \times \Delta\gamma_{\rm Bkg}(\Psi_{\rm PP}), 
\ea
where it is assumed that the CME is proportional to the magnetic field squared and the background (Bkg) is 
proportional to $v_2$.
The fraction of the measured $\Delta\gamma(\Psi_{\rm PP})$, attributable to the CME, can then be 
estimated as~\cite{Xu:2017qfs};
\ba
f_{\rm CME} = \Delta\gamma_{\rm CME}(\Psi_{\rm PP})/\Delta\gamma(\Psi_{\rm PP})   = f_{1}/f_{2},  \nonumber \\
{\rm where} \quad f_1 = \frac{r_1}{r_2} - 1\quad {\rm and} \quad  f_2 = \frac{1}{r_2^2} - 1.
\label{eq:3}
\ea

The underlying idea behind the constraints expressed in Eqs.~\ref{eq:3a}~-~\ref{eq:3} is that the $v_2$-driven background 
is  more strongly correlated with $\Psi_{\rm PP}$ [determined by the maximal particle
density in the elliptic azimuthal anisotropy and the beam axis],  than with $\Psi_{\rm RP}$ [determined 
by the impact vector $\vec{b}$ and the beam direction]. By contrast, the $\vec{B}$-field, which 
drives the CME, behaves oppositely -- weaker correlation with $\Psi_{\rm PP}$  and stronger correlation with $\Psi_{\rm RP}$. 
We will employ this new method of leveraging the measurements of $r_1$ and $r_2$ to extract 
$f_{\rm CME}$ from AMPT events as discussed below.

The operational details of the construction and the response of the $R_{\Psi_m}(\Delta S)$ correlator is
described in Refs.~\cite{Magdy:2017yje} and \cite{Magdy:2018lwk}. 
It is constructed for each  event plane $\Psi_m$, as the ratio:
\be
R_{\Psi_m}(\Delta S) = C_{\Psi_m}(\Delta S)/C_{\Psi_m}^{\perp}(\Delta S), \, m=2,3 ,
\label{eq:4}
\ee
where $C_{\Psi_m}(\Delta S)$ and $C_{\Psi_m}^{\perp}(\Delta S)$ are correlation functions
that quantify charge separation $\Delta S$, parallel and perpendicular (respectively) to 
the $\vec{B}$-field. 
$C_{\Psi_2}(\Delta S)$ measures both CME- and backgrond-driven charge separation while $C_{\Psi_2}^{\perp}(\Delta S)$ measures 
only background-driven charge separation. 
The absence of a strong correlation between the orientation of the $\Psi_3$ plane 
and the $\vec{B}$-field, also renders $C_{\Psi_3}(\Delta S)$ and $C_{\Psi_3}^{\perp}(\Delta S)$
insensitive to a CME-driven charge separation, but not to the background, so it can give crucial 
additional insight on the relative importance of background-driven and CME-driven charge separation. 
However, they are not required for the sensitivity studies presented in this work. 
%

The correlation functions used to quantify charge separation parallel to the $\vec{B}$-field, 
are constructed from the ratio of two distributions \cite{Ajitanand:2010rc}: 
\be
C_{\Psi_{m}}(\Delta S) = \frac{N_{\text{real}}(\Delta S)}{N_{\text{Shuffled}}(\Delta S)}, \, m=2,3,
\label{eq:5}
\ee
where $N_{\text{real}}(\Delta S)$ is the distribution over events, of charge separation 
relative to the $\Psi_m$ planes in each event:
\be
\Delta S = \left\langle {S_p^{h + }} \right\rangle  - \left\langle {S_n^{h - }} \right\rangle,
\label{eq:6}
\ee
%
%
\be
\Delta S = \frac{{\sum\limits_1^p {\sin (\frac{m}{2}\Delta {\varphi_{m} })} }}{p} - 
\frac{{\sum\limits_1^n {\sin (\frac{m}{2}\Delta {\varphi_{m}  })} }}{n},
\label{eq:7}
\ee
where $n$ and $p$ are the numbers of negatively- and positively charged hadrons in an event, 
$\Delta {\varphi_{m}}= \phi - \Psi_{m}$ and $\phi$ is the 
azimuthal emission angle of the charged hadrons. The $N_{\text{Shuffled}}(\Delta S)$ distribution
is similarly obtained from the same events, following random reassignment (shuffling) of the charge of 
each particle in an event. This procedure ensures identical properties for the 
numerator and the denominator in Eq.~\ref{eq:5}, except for the charge-dependent correlations 
which are of interest.
The correlation functions $C_{\Psi_{m}}^{\perp}(\Delta S)$, that quantify charge separation 
perpendicular to the $\vec{B}$-field, are constructed with the same procedure outlined 
for $C_{\Psi_{m}}(\Delta S)$, but with $\Psi_{m}$ replaced by $\Psi_{m}+\pi/m$,
to ensure that a possible CME-driven charge separation does not contribute to $C_{\Psi_{m}}^{\perp}(\Delta S)$. 

The magnitude of the CME-driven charge separation is reflected in the width $\sigma_{\Psi_2}$ of the concave-shaped 
distribution for $R_{\Psi_2}(\Delta S)$, which is also influenced by particle number 
fluctuations and the resolution of $\Psi_2$. 
That is, stronger CME-driven signals lead to narrower concave-shaped distributions (smaller widths), 
which are made broader by particle number fluctuations and poorer event-plane resolutions. 
The influence of the particle number fluctuations can be minimized by scaling $\Delta S$ 
by the width $\mathrm{\sigma_{\Delta_{Sh}}}$ of the distribution 
for $N_{\text{shuffled}}(\Delta S)$ {\em i.e.}, $\Delta S^{'} = \Delta S/\mathrm{\sigma_{\Delta_{Sh}}}$. 
%
%
\begin{figure}[t]
\includegraphics[width=0.6\linewidth, angle=-90]{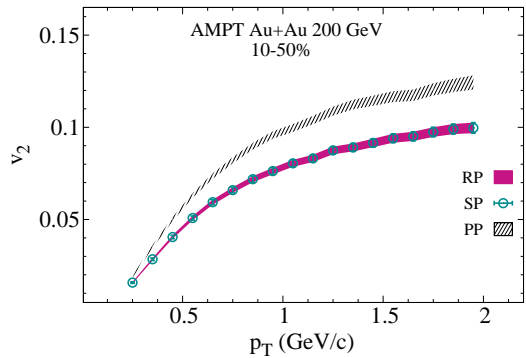}
\caption{ Comparison of the simulated $v_2(p_T)$ obtained in 10-50\% Au+Au collisions ($\sqrtsNN~=~200$ GeV) 
with $\Psi_{\rm RP}$, $\Psi_{\rm SP}$ and $\Psi_{\rm PP}$, see text.
}
\label{fig1} 
\end{figure} 
%
%
%
\begin{figure}[htb]
\includegraphics[width=0.60\linewidth, angle=-90]{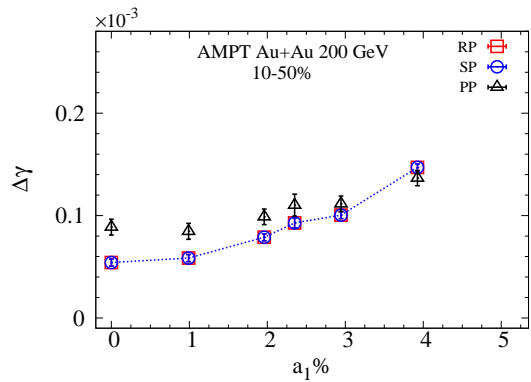}
\vskip -0.10in
\caption{ Comparison of the simulated $\Delta\gamma$ obtained in 10-50\% Au+Au collisions ($\sqrtsNN~=~200$ GeV) 
with respect to $\Psi_{\rm RP}$, $\Psi_{\rm SP}$ and $\Psi_{\rm PP}$, for several input charge separation 
fractions characterized by the $P$-odd dipole coefficient $a_1$.
} 
\label{fig2} 
\end{figure} 
%
%
\begin{figure}[!htb]
\includegraphics[width=0.60\linewidth, angle=-90]{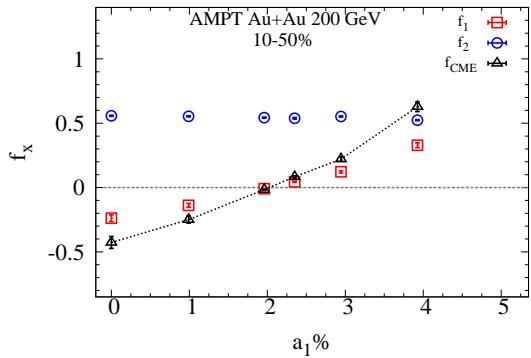}
\vskip -0.10in
\caption{ The dependence of $f_1$, $f_2$ and $f_{\rm CME}$ on different input charge separation characterized 
by the dipole coefficient $a_1$, see Eqs.~\ref{eq-f} and \ref{eq:3}.  Results are shown for 
10-50\% central Au+Au ($\sqrtsNN~=~200$ GeV) AMPT events.
} 
\label{fig3} 
\end{figure} 
Similarly, the effects of the event plane resolution can be accounted for by scaling 
$\Delta S^{'}$ by the resolution factor $\mathrm{\delta_{Res}}$, {\em i.e.}, 
$\Delta S^{''}= \Delta S^{'}\times{\delta_{Res}}$, where $\mathrm{\delta_{Res}}= \sigma_{Res}\times e^{(1-\sigma_{Res})^2}$ and 
$\mathrm{\sigma_{Res}}$ is the event plane resolution~\cite{Magdy:2017yje}. 
%
%
\begin{figure*}[!ht]
%
\includegraphics[width=0.75\linewidth, angle=-00]{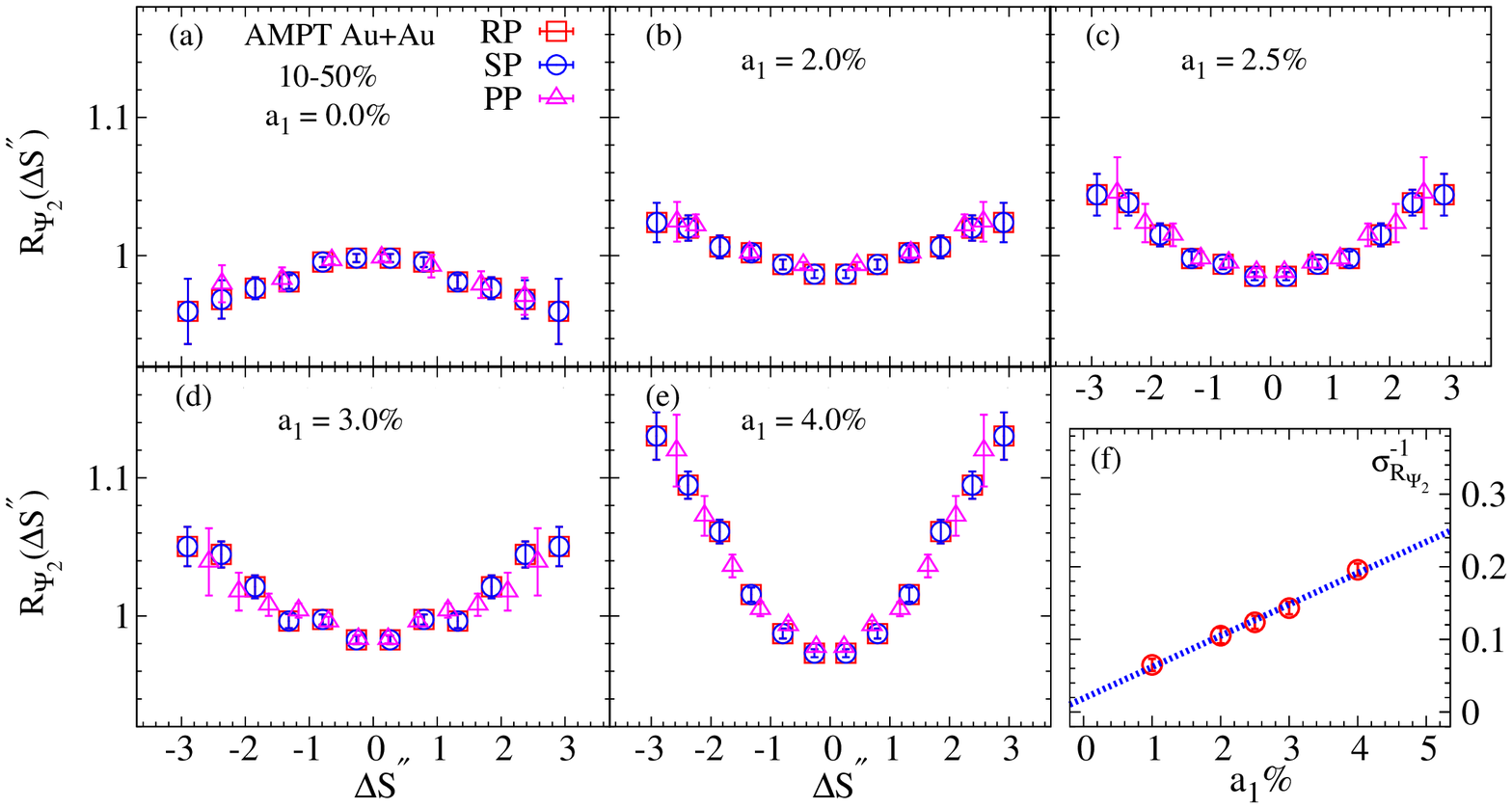}
\caption{ Comparison of the $R_{\Psi_2}(\Delta S)$ correlators obtained with respect to $\Psi_{\rm RP}$, $\Psi_{\rm SP}$  
and $\Psi_{\rm PP}$ for several $a_1$ values, as indicated, for 10-50\% Au+Au collisions at $\sqrtsNN~=~200$ GeV (a) - (e). 
Panel (f) shows the $a_1$ dependence of the inverse 
widths $\mathrm{\sigma^{-1}_{R_{\Psi_2}}}$, extracted from the  $R_{\Psi_2}(\Delta S)$ distributions; the dotted line represents 
a linear fit.
} 
\label{fig4} 
\vspace{-0.1in}
\end{figure*} 

The $10-50\%$ central AMPT events, generated for several input values of charge separation $f_{0}$ (cf. Eq.~\ref{eq-f}), 
relative to the reaction- $\Psi_{\rm RP}$, spectator- $\Psi_{\rm SP}$ and the participant plane $\Psi_{\rm PP}$, were 
analyzed to extract $f_{\rm CME}$ via the $\Delta\gamma$ correlator and $\sigma_{\Psi_2}$
via the $R_{\Psi_2}(\Delta S)$ correlator.  Approximately $10^6$ events were generated for each value of $f_{0}$.
The analyses included charged particles with $|\eta| < 1.0$
and transverse momentum $0.2 < p_T < 2$~GeV/$c$. To enhance the statistical significance of the 
measurements,  the participant plane $\Psi_{\rm PP}$  was determined with charged hadrons in the 
range $2.5< \eta <4.0$ . The charge separation of charged hadrons in $|\eta| < 1.0$ 
were then measured relative to $\Psi_{\rm PP}$.
 Representative results are summarized in Figs.~\ref{fig1}~-~\ref{fig4}. 

Figure \ref{fig1} compares the $v_2(p_T)$ obtained with $\Psi_{\rm RP}$, $\Psi_{\rm SP}$ and $\Psi_{\rm PP}$ for
10-50\% Au+Au collisions.
It shows the expected similarity between the results for $\Psi_{\rm RP}$ and $\Psi_{\rm SP}$, as well as larger values 
for $\Psi_{\rm PP}$ that confirm the enhanced fluctuations associated with the participant geometry and consequently,
the initial-state eccentricity $\varepsilon_2$. 
This difference is essential for the procedure outlined in Eqs.~\ref{eq:3a} - \ref{eq:3}. 

A similar comparison of the $\Delta\gamma$ results for the three planes 
is given in Fig.~\ref{fig2}.
It shows that for $a_1 \alt 3\%$, the $\Delta\gamma$ values obtained with $\Psi_{\rm PP}$ are larger than those
obtained with $\Psi_{\rm RP}$ and $\Psi_{\rm SP}$; there is also little, if any, difference between the values  obtained 
with  $\Psi_{\rm RP}$ and $\Psi_{\rm SP}$ over the full range of the input $a_1$ values. This latter trend is to be expected since 
the fluctuation of $\Psi_{\rm SP}$ about $\Psi_{\rm RP}$ is small.
For $a_1 \agt 4\%$, the $\Delta\gamma$ values for $\Psi_{\rm RP}$ and $\Psi_{\rm SP}$ 
become larger than the ones for $\Psi_{\rm PP}$ (not shown in Fig.~\ref{fig2}), consistent with 
a stronger influence from the proxy CME-driven charge separation.

The extracted values  of $v_{2}$ and $\Delta\gamma$, with respect to $\Psi_{\rm PP}$ and 
$\Psi_{\rm SP}$ were used to evaluate $f_1$, $f_2$ and $f_{\rm CME}$ following the procedure 
outlined in Eqs.~\ref{eq:3a} - \ref{eq:3}~\cite{Xu:2017qfs}. Fig.~\ref{fig3} summarizes the $a_1$ dependence of 
$f_1$, $f_2$ and $f_{\rm CME}$. It indicates a flat $f_2$, consistent with the expectation 
that the $v_2$ fluctuations should be relatively insensitive to the introduction of small  $a_1$ signals.
By contrast, $f_1$ and $f_{\rm CME}$, which are both negative for $a_1 \alt 2.5\%$, show an 
increase with $a_1$ and become positive for $a_1 \agt 2.5\%$.  The negative values observed for
$f_{\rm CME}$ suggests that for $a_1 \alt 2.5\%$, the correlator is either (i) unable to make the robust 
distinction between signal and background required to measure the input proxy CME-signal
or (ii) the assumptions used to estimate $f_{\rm CME}$ are invalid.
Note that $f_{\rm CME} $ is only 0.6 even for a relatively large input signal of $a_1 = 4.0\%$.

The change from negative to positive values for $f_{\rm CME}$ (cf. Fig.~\ref{fig3})
suggests a {\em ``turn-on''} $a_1$ value, below which, the modified $\Delta\gamma$ 
correlator (cf. Eqs.~\ref{eq:3a} - \ref{eq:3}) is unable to detect a CME-driven signal. 
This detection threshold could pose a significant limitation for CME detection and characterization 
with this correlator, because it is comparable to, or larger than the magnitude of the CME-driven 
charge separation expected in actual experiments. Equally important is the fact that  $f_{\rm CME} \le 0.0$ does 
not give a robust indication of the absence of a CME signal. The latter could have important implications 
for the interpretation of current and future  $f_{\rm CME}$ measurements.

The sensitivity of the $R_{\Psi_{\rm XX}}(\Delta S)$ (${\rm {\small XX}=RP, SP, PP}$) correlator to varying 
degrees of input CME-driven charge  separation (characterized by $\mathrm{a_{1}}$) was studied 
using the same AMPT events employed 
in the leveraged $\Delta\gamma$ study. Figs.~\ref{fig4}(a) - (e) show the $R_{\Psi_{\rm XX}}(\Delta S)$ correlator 
distributions obtained for $10-50\%$ central Au+Au collisions, relative to $\Psi_{\rm RP}$, $\Psi_{\rm SP}$ 
and $\Psi_{\rm PP}$ for several values of $a_1$ as indicated. In  each of these plots, $\Delta S$
is scaled to account for the effects of number fluctuations and event plane 
resolution as outlined earlier and in Ref.~\cite{Magdy:2017yje}. 

The concave-shaped distribution, apparent in each panel of Fig.~\ref{fig4} (b) - (e), confirms the input 
charge separation signal in each case; note the weakly convex-shaped distribution for $a_1=0$ in Fig.~\ref{fig4} (a). 
Note as well that in contrast to the $\Delta\gamma$ correlator,
the $R_{\Psi_{\rm XX}}(\Delta S)$ distributions are independent of the plane used to measure them,
suggesting that they are less sensitive to the $v_2$ driven background and their associated fluctuations.
The apparent decrease in the widths of these distributions with $a_1$, also confirm the expected trend.  

To quantify the implied signal strengths, we extracted the width $\mathrm{\sigma_{R_{\Psi_2}}}$ of the 
$R_{\Psi_{2}}(\Delta S)$ distributions obtained for the respective values of $a_1$.  
Fig.~\ref{fig4}(f) shows the inverse widths $\mathrm{\sigma^{-1}_{R_{\Psi_2}}}$ vs. $a_1$.
They indicate an essentially linear dependence on $a_1$ (note the dotted line fit). 
Here, it is noteworthy that for $a_1 \alt 0.5\%$, significant additional statistics are required to determine 
 $\mathrm{\sigma_{R_{\Psi_2}}}$ with good accuracy. These results suggests that the 
$R_{\Psi_{m}}(\Delta S)$ correlator not only suppresses background, but is sensitive to very small CME-driven 
charge separation in the presence of such backgrounds.
 
In summary,  we have used both the $R_{\Psi_{2}}(\Delta S)$ correlator and an event-plane-leveraged
version of the $\Delta\gamma$ correlator to analyze AMPT events with varying degrees of input 
proxy CME signals.
Our sensitivity study indicates a {\em turn-on} threshold for $f_{\rm CME}=\Delta\gamma_{\rm CME}/\Delta\gamma$, 
which renders the leveraged $\Delta\gamma$-correlator insensitive to input signals with
$a_1 \alt 2.5\%$. The magnitude of this detection threshold,  which is comparable to that for the 
purported signal in heavy ion collisions and less than
the signal difference for isobaric collisions,  could pose significant restrictions on 
its use to detect the CME. By contrast, the $a_1$-dependent $R_{\Psi_{2}}(\Delta{S})$ 
correlators indicate inverse widths $\mathrm{\sigma^{-1}_{R_{\Psi_2}}}$, that are linearly 
dependent on $a_1$, and independent of the character of the 
event plane ($\Psi_{\rm RP}$, $\Psi_{\rm SP}$ or $\Psi_{\rm PP}$) used  
for their extraction. These results not only have implications for the interpretation  of 
current and future $f_{\rm CME}=\Delta\gamma_{\rm CME}/\Delta\gamma$ measurements; they further 
indicate that the $R_{\Psi_{2}}(\Delta{S})$ correlator can provide robust quantification of minimal CME-driven 
charge separation in the presence of realistic backgrounds, that could aid characterization 
of the CME in RHIC and LHC collisions.

\section*{Acknowledgments}
\begin{acknowledgments}
This research is supported by the US Department of Energy, Office of Science, Office of Nuclear Physics, 
under contracts DE-FG02-87ER40331.A008  (RL), DE-FG02-94ER40865 (NM) and by the 
National Natural Science Foundation of China under Grants No. 11890714, No. 11835002, 
No. 11961131011, and No. 11421505, the Key Research Program of the Chinese Academy 
of Sciences under Grant No. XDPB09.
\end{acknowledgments}
%
%
\bibliography{lpvpub} 
\end{document}